\documentclass[aps,prl,twocolumn,showpacs,superscriptaddress,groupedaddress]{revtex4}  
\usepackage{graphicx}  
\usepackage{dcolumn}   

\usepackage{bm}        
\usepackage{amssymb}   

\def\fiverm{\footnotesize}
\input prepictex.tex
\input pictex.tex
\input postpictex.tex

\usepackage[usenames,dvipsnames,svgnames]{xcolor}

\def\2;{\;\;}

\def\Sfrac#1#2{\hbox{\large $\frac{#1}{#2}$}}
\def\sfrac#1#2{\hbox{\nor $\frac{#1}{#2}$}}

\def\LB{\left(}         \def\RB{\right)}

\def\nor{\normalsize}

\def\Tiny#1{\scalebox{0.5}{#1}}

\definecolor{blue}{rgb}{0,0.18,0.39}
\definecolor{RoyalBlue}{rgb}{0,0.2,0.7}

\def\axes#1#2#3#4#5#6#7{
\setplotarea x from #7 to #5, y from #2 to #6
\setplotarea x from #1 to #5, y from #2 to #6
\axis left shiftedto x=#3 
 /
\axis bottom shiftedto y=#4
/
\put {\footnotesize$\bullet$} at #3 #4
}

\definecolor{Maroon}{cmyk}{0,0.87,0.68,0.62}
\definecolor{Brown}{rgb}{0.7,0.3,0}
\definecolor{Navy}{rgb}{0.3,0.0,0.4}
\definecolor{Red}{cmyk}{0,1,1,0}
\definecolor{BrickRed}{cmyk}{0.16,0.89,0.61,0.02}
\definecolor{DarkRed}{cmyk}{0,1,1,0.5}
\definecolor{DarkBlue}{cmyk}{1,1,0,0.2}
\definecolor{DarkGreen}{cmyk}{1,0,1,0.4}
\definecolor{Green}{cmyk}{1,0,1,0}
\definecolor{DarkBrown}{cmyk}{0,0.81,1,0.6}
\definecolor{OrangeRed}{cmyk}{0,1,0.87,0}
\definecolor{RedOrange}{cmyk}{0,0.77,0.87,0}
\definecolor{Orange}{cmyk}{0,0.61,0.87,0}
\definecolor{Offwhite}{rgb}{.8,0.9,.8}
\definecolor{Offwhite2}{cmyk}{.04,.02,.01,0}
\definecolor{Tan}{rgb}{0.82,0.70,0.55}
\definecolor{Blue}{rgb}{0,0,1}
\definecolor{RoyalBlue}{rgb}{0.25,0.41,0.88}
\definecolor{Sepia}{rgb}{0.37,0.14,0.07}
\definecolor{myblue}{cmyk}{0.025,0.05,0,0}
\definecolor{Mahogany}{cmyk}{0.18,0.87,1,0.08}

\definecolor{green1}{cmyk}{0.25,0,0.76,0}
\definecolor{green2}{cmyk}{0.25,0,0.76,0.07}
\definecolor{green3}{cmyk}{0.25,0,0.76,0.20}
\definecolor{green4}{cmyk}{0.25,0,0.75,0.30}
\definecolor{green5}{cmyk}{0.25,0,0.75,0.40}
\definecolor{green6}{cmyk}{0.25,0,0.75,0.50}

\definecolor{B02}{cmyk}{0,0.14,0.22,0.12}
\definecolor{B03}{cmyk}{0,0.16,0.26,0.16}
\definecolor{B04}{cmyk}{0,0.19,0.28,0.19}
\definecolor{B05}{cmyk}{0,0.25,0.32,0.25}
\definecolor{B06}{cmyk}{0,0.31,0.36,0.31}
\definecolor{B07}{cmyk}{0,0.37,0.40,0.37}
\definecolor{B08}{cmyk}{0,0.46,0.46,0.46}
\definecolor{B09}{cmyk}{0,0.55,0.52,0.54}
\definecolor{B10}{cmyk}{0,0.69,0.61,0.62}
\definecolor{B11}{cmyk}{0,0.78,0.70,0.68}
\definecolor{B12}{cmyk}{0,0.93,0.85,0.60}
\definecolor{B13}{cmyk}{0.5,1,0.6,0.30}
\definecolor{B14}{cmyk}{1,1,0,0.30}
\definecolor{B15}{cmyk}{1,1,0,0}

\begin{document}

\widetext
\leftline{Version 6.0 as of \today}

\title{Osmotic pressure of compressed lattice knots}
\author{EJ Janse van Rensburg}
\address{Department of Mathematics \& Statistics, York University, Toronto, ON, Canada, M3J 1P3}
\date{\today}

\begin{abstract}
A numerical simulation shows that the osmotic pressure of compressed lattice
knots is a function of knot type, and so of entanglements.  The osmotic pressure
for the unknot goes through a negative minimum at low concentrations, but
in the case of non-trivial knot types $3_1$ and $4_1$ it is negative for low concentrations.
At high concentrations the osmotic pressure is divergent, as predicted by
Flory-Huggins theory.  The numerical results show that each knot type has an
equilibrium length where the osmotic pressure for monomers to migrate into
and out of the lattice knot is zero. Moreover, the lattice unknot is found to have two
equilibria, one unstable, and one stable, whereas the lattice knots of type 
$3_1$ and $4_1$ have one stable equilibrium each.
\end{abstract}

\pacs{36.20.-r,61.25.Hq, 82.35.-x,87.15.-v}
\maketitle


\section{Introduction}

Confinement of a biopolymer (eg. in an organel in a living cell) causes an 
increase in knotting (and so in topological entanglement) \cite{Orlandini}.  The
confinement compresses the biopolymer and this is known to increase 
self-entanglements in the backbone of the polymer \cite{Tang}.  
It is known that entanglements have an effect on the physical properties
of biopolymer such as DNA \cite{Cozzarelli}. Topological
entanglement (knotting and linking) also changes the movement of DNA, 
for example in the ejection of DNA from a viral capsid \cite{Marenduzzo}
or the speed of electrophoretic migration of polymers 
\cite{Lim1993,Stasiuk1996,Weber06}.

\begin{figure}
\input{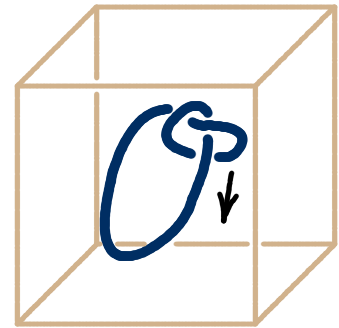}
\caption{A model of a knotted ring polymer in a cavity.  Contributions to the
entropy are due to translational degrees of freedom, topological constraints
due to the knot, and conformational degrees of freedom.}
\label{figure1}
\end{figure}

In this paper a model  of a confined ring polymer is examined.   A particular 
simple model is to place a random circular string 
\cite{Flory56,Edwards65,Edwards67,Edwards68} in a box (see Fig.~\ref{figure1}).  
Conformational entropy of the string is reduced by the confining environment 
of the box.  In addition, entanglements can be modelled and controlled by
fixing the knot type of the string.  The entropy of the confined string can be
quantified by placing it in a lattice.  If it is self-avoiding, then it is a closed self-avoiding
walk \cite{Flory56,Hammersley57,Flory69} and this is a lattice model of
a ring polymer in a good solvent.  

While the model in Fig.~\ref{figure1} is interesting from a purely theoretical perspective, it 
can also be seen as a very simplified model giving a qualitative understanding of the role of 
entanglement in the properties of random string-like objects, such as, for example,
DNA and other biopolymers. DNA is a double helix linear polymer normally 
compressed and compacted in small volumes \cite{Cairns63}.  Enzymes 
unwind and release segments of DNA to mediate cellular processes.  
These released segments have increased conformational degrees of freedom 
while also being entangled and  connected to and confined by other structures in the cell.  
The entanglements and confining environment reduce the conformational entropy of 
these segments.  Such segments are also subject to random mutations by random events,
which may cause deletion or insertion of basepairs (or even sequences of basepairs) 
in the genome \cite{Clancy08,Berg80,Kunkel00}.  This changes the length of the DNA and
the tendency to gain or reduce the length of the DNA backbone can be modelled as 
an osmotic pressure of basepairs in the backbone.   In another situation, but 
also involving DNA, the length of segments of DNA are changed by intercalating drugs 
\cite{Camunas2015}.  In vitro these drugs are in equilibrium in a solvent or bound
to the DNA backbone, and so the model in Fig.~\ref{figure1} can similarly be seen as
a simplified model of the osmotic pressure of intercalating drugs bound to the DNA backbone.

\section {Compressed lattice knots}

 A closed self-avoiding walk is
a \textit{lattice polygon}  \cite{Hammersley61,Frisch61,Delbruck62}.
Lattice polygons are knotted asymptotically with probability $1$ 
\cite{Sumners88,Pippenger89}. A lattice polygon with fixed knot type is
a \textit{lattice knot} \cite{DeGennes84,Diao1993}.  It is known that
the entropy of a lattice knot is a function of its knot type \cite{Orlandini1998,JvR2014}.  
\textit{Tight lattice knots} \cite{DeGennes84} are minimal length lattice
knots \cite{JvR1995,JvR2011,Scharein2009}.  The compressibility of tight
lattice knots is known to be a function of knot type \cite{JvR2012,Gasumova2012}.

\begin{figure}
\input{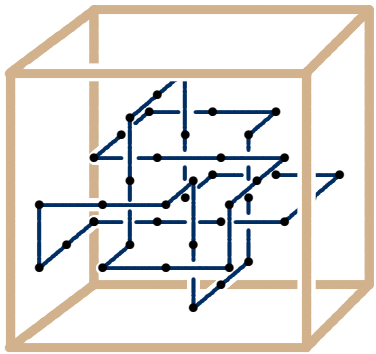}
\caption{A lattice knot in a cubical cavity.}
\label{figure2}
\end{figure}

\begin{figure}
\input{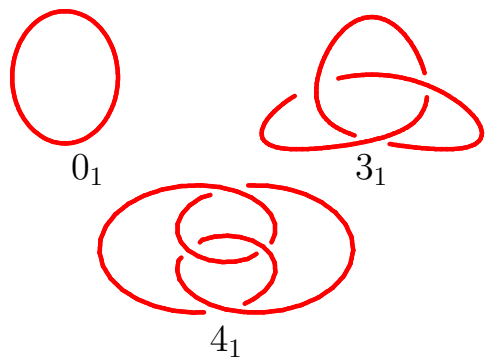}
\caption{\label{figure3} The unknot $0_1$, the
trefoil knot $3_1$ and the figure eight knot $4_1$ \cite{Adams04,Rolfsen}.}
\end{figure}

In Fig. \ref{figure2} the model in Fig. \ref{figure1} is quantified by a placing
and compressing a lattice knot in a cubical box (this is a lattice version 
with self-avoidance of the model in reference \cite{Arsuaga04}).  The entropy
of the \textit{compressed lattice knot} has contributions from translational degrees 
of freedom (if it is small compared to the side length of the box), from topological 
constraints (due to entanglements which depends on the knot type), and from conformational
degrees of freedom.  In this paper compressed lattice knots of 
three knot types \cite{Adams04,Rolfsen}, namely the \textit{unknot} ($0_1$), 
the \textit{trefoil} ($3_1$), and the \textit{figure eight knot} ($4_1$) will be considered;
see Fig. \ref{figure3}. 

A cube in the lattice of side-length $L{-}1$ has \textit{volume} $V=L^3$ and
\textit{side-length} $L$.  The maximum length of a lattice knot confined 
to a cube of side-length $L$ is $L^3$ if $L$ is even, and $L^3{-}1$ if $L$ is odd.  
The \textit{lattice unknot} has minimal length $4$ and there are $3L(L{-}1)^2$ 
ways it can be placed in the cube.  The \textit{lattice trefoil knot} $3_1$ can be 
tied with $24$ steps in the cubic lattice \cite{Diao1993}, and there 
are $3328$ conformations distinct under translations in the cubic lattice 
\cite{Scharein2012}.  None of these tight lattice trefoils can be realised in a
cube of side-length $3$, but a numerical simulation detected
$4168$ distinct placements of $3304$ tight lattice trefoils 
in a cube of side-length $4$, and $30104$ 
distinct placements of tight lattice trefoils in a cube of side-length
$5$.  Similarly, a \textit{tight lattice figure eight knot} $4_1$ has minimal
length $30$ in the cubic lattice \cite{Scharein2012} and there are $3648$
conformations distinct under translations in the cubic lattice.  A computer
count shows that none of these can be realised in a cube of side-length $3$, 
but there are $864$ distinct placements of tight lattice figure eight knots
in a cube of side-length $4$, and $18048$ distinct placements in a cube of side-length $5$. 

\section{Free energy and osmotic pressure}

Denote the number of distinct placements of lattice knots of length $n$, 
of knot type $K$, confined in a cube of side-length $L$, by $p_{n,L}(K)$.  
Then, for example, $p_{24,3}(3_1)=0$ and $p_{24,4}(3_1)=4168$.  
Approximate enumeration of $p_{n,L}(K)$ can be done by using the 
GAS algorithm \cite{JvR2009} implemented with BFACF 
moves \cite{BFACF1,BFACF2}.  See reference \cite{JvR2011A} for details.

The concentration of vertices in a lattice knot in a cube
of side-length $L$ is $\phi=\sfrac{n}{V}$ where $V=L^3$.  The free energy 
at concentration $\phi$ of lattice knots of type $K$ is
\begin{equation}
F_{tot} (\phi;K) = - \log p_{n,L} (K) ,
\end{equation}
where $n=\phi\,V$. 
The free energy per unit volume is $F_L(\phi;K) = \sfrac{1}{V} F_{tot} (\phi;K)$
and this is plotted in Fig. \ref{figure4} for $2\leq L \leq 15$ and for $K=0_1$
(the unknot) against the monomer concentration $\phi$.  The shape of
these curves is consistent with prediction of Flory-Huggins theory \cite{DeGennes79}.

\begin{figure}
\input{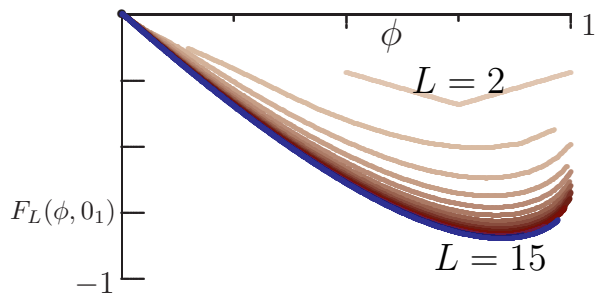}
\caption{\label{figure4} The free energy per unit volume for unknotted
lattice knots, for $2\leq L \leq 15$.}
\label{figure4}
\end{figure}

\begin{figure}[h!]
\input{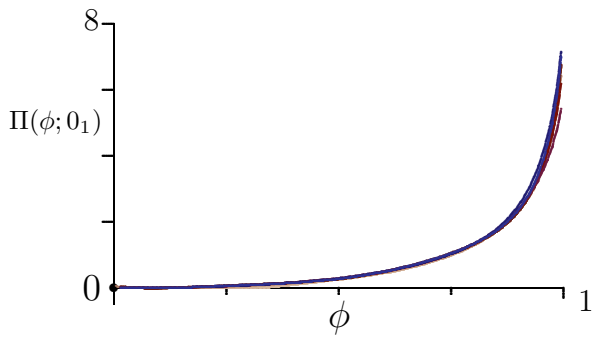}
\caption{\label{figure5} The calculated osmotic pressure of compressed
lattice unknots.  The data are for $2\leq L \leq 15$.}
\end{figure}

\begin{figure}
\input{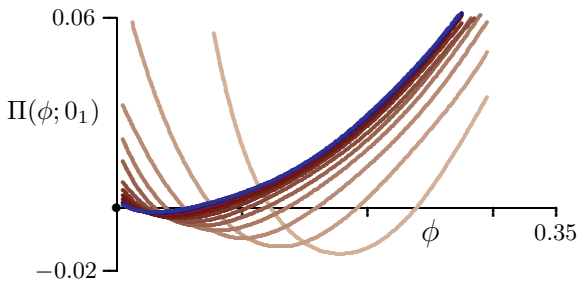}
\caption{\label{figure6} The osmotic pressure of compressed lattice 
unknots at low concentration for $3\leq L \leq 15$.}
\end{figure}

The osmotic pressure $\Pi(\phi;K)$ of compressed lattice knots is given by 
\begin{equation}
\Pi(\phi;K) = - \Sfrac{d}{dV}\, F_{tot}(\phi;K) .
\end{equation}
Changing variables to $\phi$ shows that 
\begin{equation}
\Pi(\phi;K) = \phi^2 \, \Sfrac{d}{d\phi} \LB \sfrac{1}{\phi} \, F_L(\phi;K) \RB
\end{equation}
in terms of the free energy per unit volume.  This can be computed from the data
in figure \ref{figure4} by taking a numerical derivative.  Using a central second
order numerical approximation to the derivative gives Fig. \ref{figure5}. 
This appears to be consistent with the predicted Flory-Huggins osmotic
pressure:  $\Pi(\phi;0_1)$ is increasing and sharply diverges as 
$\phi\to 1^-$.  Closer examination of Fig. \ref{figure5} shows that 
$\Pi(\phi;0_1)$ is not monotone but is decreasing a low concentrations
and negative and not monotonic on an interval of low concentrations; 
see Fig. \ref{figure6}. 

\begin{figure}
\input{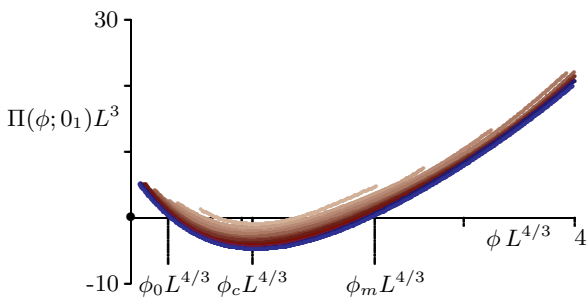}
\caption{\label{figure7} Rescaled osmotic pressures for the unknot 
$0_1$, plotted as a function of $\phi\,L^{4/3}$.  The data are for
$3\leq L \leq 15$.}
\end{figure}

At negative osmotic pressure the lattice unknot will add length.   Similarly, at 
positive osmotic pressure the lattice unknot will shed length and become smaller.
The pressure curves in Fig. \ref{figure6} are functions of $L$ and each has two zeros at 
$\phi_0$ and $\phi_m$.  At concentrations $\phi<\phi_0$ the lattice unknot will 
evaporate, and when $\phi_0<\phi<\phi_m$ it will add length until the concentration
is $\phi_m$ (which is a stable fixed point).  It will also shed length if $\phi>\phi_m$
until the concentration is $\phi_m$.

Lattice polygons of length $n$ has linear size $O(n^\nu)$ where $\nu\approx
\sfrac{3}{5}$ is the metric exponent in three dimensions \cite{Flory69} (a more
accurate estimate is $\nu \approx 0.587597(7)$ \cite{Clisby10}).  Effects of the 
confining cube will become important when $n^\nu \sim L$.  The osmotic pressure
should vanish at this point;  the result is that $\phi_m \sim L^{1/\nu}/L^3 =
L^{1/\nu-3}$.  As $\phi\to 0^+$,  $\Pi(\phi;K) \sim L^{-3}$.  Using the Flory value 
for $\nu$ and then plotting $\Pi(\phi;K)\,L^3$ as a function of 
$\phi\,L^{3-1/\nu} \approx \phi\,L^{4/3}$ should collapse the data in 
Fig. \ref{figure6}.  This is shown in Fig. \ref{figure7}, although there are 
still finite size corrections.  Extrapolating the zeros of the curves in 
Fig. \ref{figure7} gives $\phi_0 \simeq 0.149\, L^{-4/3}$,
$\phi_m \simeq 0.286\, L^{-4/3}$.  Since the osmotic pressure vanishes at
these concentrations the equilibrium lengths at which the osmotic pressure 
vanishes are $n_0 \simeq 0.149\, L^{5/3}$ and $n_m \simeq 0.286\, L^{5/3}$.  
The osmotic pressure of the unknot goes throuh a minimum at 
$\phi_c \simeq 0.209\, L^{-4/3}$.

\begin{figure}
\input{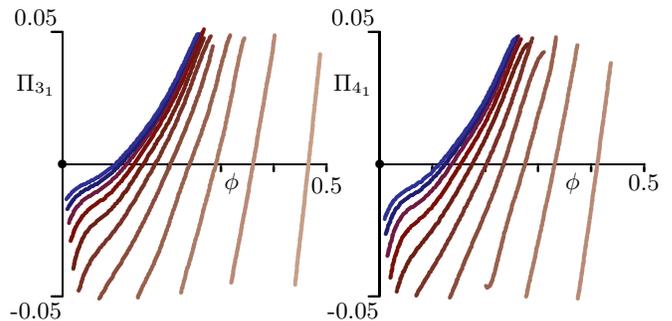}
\caption{\label{figure8} The osmotic pressure $\Pi_{3_1} \equiv \Pi(\phi;3_1)$ of the
trefoil knot, and $\Pi_{4_1} \equiv \Pi(\phi;4_1)$ of the figure eight knot plotted
against the concentration $\phi$ for $0\leq \phi \leq 0.5$.}
\end{figure}

\begin{figure}
\input{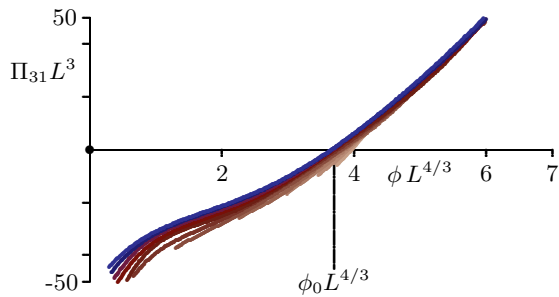}
\caption{\label{figure9} Rescaled osmotic pressures
$\Pi_{3_1} \equiv \Pi(\phi;3_1)$ for lattice knots of type
$3_1$ (trefoil).  The data are taken from the left panel in Fig.
\ref{figure8} for $4\leq L \leq 15$.}
\end{figure}

\begin{figure}
\input{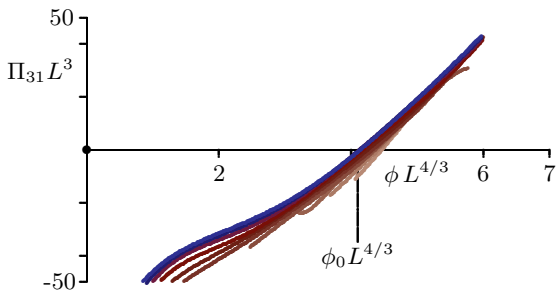}
\caption{\label{figure10} Rescaled osmotic pressures 
$\Pi_{4_1} \equiv \Pi(\phi;4_1)$ for lattice knots of type
$4_1$ (figure eight knot).  The data are taken from the
right panel in Fig. \ref{figure8} for $4\leq L \leq 15$.}
\end{figure}

The osmotic pressures of compressed lattice knots at low concectration
and of knot types $3_1$ and $4_1$ are plotted in Fig. \ref{figure8}.  Here
the osmotic pressures are monotone increasing with concentration $\phi$, passing 
through zero at a critical concentration $\phi_0$. Rescaling the data in the same way 
as in Fig.  \ref{figure7} gives Figs. \ref{figure9} and \ref{figure10}.  This shows that
for $3_1$, $\phi_0\, L^{4/3} \simeq 3.94\, L^{-4/3}$, and for $4_1$,
$\phi_0\, L^{4/3} \simeq 4.48\, L^{-4/3}$. For $\phi<\phi_0$ the osmotic
pressure is negative and the lattice knot will grow to an equilibrium length
$n_0 \simeq 3.94\, L^{5/3}$ for $3_1$ and $n_0 \simeq 4.48\, L^{5/3}$ for $4_1$.  

\section{Conclusion} 

In this letter a numerical simulation of compressed lattice knots as a model
of an entangled ring polyer show that the osmotic pressure is a function 
of knot type.  Since the level of entanglements is a function of knot type,
these results support the notion that the properties of confined 
biopolymers, such as DNA, is a function of the level of entanglement 
if the biopolymer is confined or compressed in a narrow space, or adsorbed 
on a membrane.  Adsorption of the knotted polymer on the surface of a membrane was
analised in the lattice in reference \cite{Vanderzande}.  If the polymer can relax
freely after adsorption, then the knot localizes and its effects dissappear 
as the length of the polymer increases \cite{Metzler02,Ercolini07}.  
On the other hand, the adsorbed polymer should have properties of projected 
three dimensional polymers;  experimental evidence of this was given in 
reference \cite{Ercolini07}.

The numerical data in this paper show that the rescaled osmotic pressure vanishes
at a critical concentrations, as shown in Figs. \ref{figure7},
\ref{figure9} and \ref{figure10}, and that these critical concentrations are
functions of knot types.  In the case of the unknot there are two critical concentrations
where the osmotic pressure vanishes.  At these concentrations the lattice unknot has an 
equilibrium length, but at the lower critical concentration this is unstable, and the unknot
will tend to grow or evaporate.  At the higher critical concentration the equilibrium length
is stable.  The situation is not the same for the trefoil and figure eight knot types.
In these cases there is one stable equilibrium at a critical concentration which
is dependent on knot type.

\section*{Acknowledgements}
EJJvR acknowledges support from NSERC(Canada) in the form of Discovery Grant
RGPIN-2014-04731.

\vfill\eject


\end{document}